\documentclass[aps,pre,twocolumn,superscriptaddress,groupedaddress,showpacs]{revtex4}
\usepackage{graphicx,epsf,dsfont,amssymb,verbatim}

\RequirePackage{color}

\begin{document}
\title{Entropic equation of state and scaling functions near the critical point in scale-free
networks}
\author{C.\
von Ferber} \email[]{C.vonFerber@coventry.ac.uk}
 \affiliation{Applied Mathematics Research Centre, Coventry University,
Coventry CV1 5FB, UK}
 \affiliation{Physikalisches Institut
Universit\"at Freiburg, D-79104 Freiburg, Germany}
\author{R. Folk}\email[]{reinhard.folk@jku.at}
 \affiliation{Institut f\"ur Theoretische Physik, Johannes Kepler
Universit\"at Linz, A-4040, Linz, Austria}
\author{Yu. Holovatch}\email[]{hol@icmp.lviv.ua}
\affiliation{Institute for Condensed Matter Physics, National
Academy of Sciences of Ukraine, UA--79011 Lviv, Ukraine}
\affiliation{Institut f\"ur Theoretische Physik, Johannes Kepler
Universit\"at Linz, A-4040, Linz, Austria}
\author{R. Kenna}\email[]{R.Kenna@coventry.ac.uk}
\affiliation{Applied Mathematics Research Centre, Coventry
University, Coventry CV1 5FB, UK}
\author{V. Palchykov}\email[]{palchykov@icmp.lviv.ua}
\affiliation{Institute for Condensed Matter Physics, National
Academy of Sciences of Ukraine, UA--79011 Lviv, Ukraine}
\begin{abstract}
We analyze the entropic equation of state for a many-particle
interacting system in a scale-free network. The analysis is
performed in terms of scaling functions which  are of fundamental
interest in the theory of critical phenomena and have previously
been theoretically and experimentally explored in the context of
various magnetic, fluid, and superconducting systems in two and
three dimensions. Here, we obtain general scaling functions for the
entropy, the constant-field heat capacity, and the isothermal
magnetocaloric coefficient near the critical point in scale-free
networks, where the node-degree distribution exponent $\lambda$
appears to be a global variable and plays a crucial role, similar to
the dimensionality $d$ for systems on lattices. This extends the
principle of universality to systems on scale-free networks and
allows quantification of the impact of fluctuations in the network
structure on critical behavior.
\end{abstract}

\date{January 19, 2011}
\maketitle

\section{Introduction \label{I}}

Phase transitions and critical behavior in complex networks currently attract
 much attention \cite{Dorogovtsev08} because of their
unusual features and broad array of applications, ranging from
socio- \cite{sociophysics} to nanophysics \cite{Tadic05}. It is by
now well established that the critical behavior of   a many-particle
interacting system located on the nodes of a general network may
crucially differ from that of a system located on the sites of a
$d$-dimensional regular lattice. Of particular interest are the
so-called scale-free networks, for which the probability to find a
node of degree $k$ (i.e. with $k$ nearest neighbors) vanishes for
large $k$ as a power law

\begin{equation} \label{1.1}
P(k) \sim k^{-\lambda}.
\end{equation}

The questions we address in this paper concern two fundamental
principles of critical phenomena: universality and scaling
\cite{Stanley_Domb}. Both of these questions have to be reconsidered when a
system resides on a network. Usually, the universality of critical
phenomena is understood as stating that the thermodynamic
properties near the critical point $T_c$ are governed by a small
number of global features, such as dimensionality, symmetry, and the type of
interaction. In turn, the scaling hypothesis states that the
singular part of a thermodynamic potential near $T_c$ has the form
of a generalized homogeneous function. To be specific, for the
Helmholtz potential of a magnetic system the latter can be written
as \cite{note1}
\begin{equation}\label{1.2}
F(\tau,M) = \tau^{2-\alpha} f_{\pm}(M/\tau^{\beta}),
\end{equation}
where $M$ is the magnetization, $\tau=|T-T_c|/T_c$, $\alpha$,
$\beta$ are the universal exponents, and the sign $\pm$ corresponds
to $T>T_c$ or $T<T_c$ respectively. The essence of relation
(\ref{1.2}) is that the two-variable function $F(\tau,M)$, when
appropriately rescaled, is expressed in terms of a single scaling
variable leading to the {\em scaling function} $f_\pm(x)$. The
expression (\ref{1.2}) gives an example of the scaling function for
the Helmholtz potential $F(\tau,M)$. Together with other scaling functions - for
the equation of state and for thermodynamic functions - this appears to
give a suitable and comprehensive description of critical phenomena
\cite{scaling_functions,Privman91}. These scaling functions are
also universal in the sense explained above.

For systems on scale-free networks the principle of universality is
extended: there, the node-degree distribution exponent $\lambda$ in
Eq.(\ref{1.1}) appears to be a global variable and plays a crucial
role, similar to the dimension $d$ for systems on lattices (see e.g.
\cite{Leone02,Dorogovtsev02,Goltsev03,Igloi02,Palchykov09,Palchykov10}).
All systems that belong to a given universality class are governed
by the same values of critical exponents and critical amplitude
ratios and share the same universal form of scaling functions.
Recently, the scaling function formalism has been applied to
describe the critical behavior of magnetic systems with the
structure of a scale-free network \cite{Palchykov10}. There, scaling
functions for the magnetic equation of state and isothermal
susceptibility were derived. In this paper we are interested in the
entropic form of the equation of state. In particular this opens the
possibility to derive scaling functions for the heat capacity. These
are of wide and fundamental interest in the theory of critical
phenomena and have been the subject of thorough theoretical and
experimental studies for various magnetic, fluid, and
superconducting systems
\cite{Krasnow73,Simons74,Barmatz75,Aharony76,Salamon88,Stroka92,%
Plackowski05,Franco09}. A particular point of interest concerns the
existence or non-existence of the moments of the distribution
(\ref{1.1}) on the universal behavior.

 The set-up of the the paper is as
follows: in section \ref{II} we define the notations and derive
expressions for the entropic equation of state, heat capacity and
magnetocaloric coefficient scaling functions for systems on
scale-free networks. These expressions are further analyzed and
compared with corresponding functions for bulk systems in section
\ref{III}. The paper concludes with a summary and an outlook in section
\ref{IV}.

\section{Free energy and scaling functions \label{II}}

The critical behavior of a many-particle system interacting on a
scale-free network crucially depends on the
node degree distribution $P(k)$ via the decay exponent $\lambda$ in Eq.(\ref{1.1}) \cite{Dorogovtsev08}. In particular, for an infinite network the value of $\lambda$
determines the order of the first diverging moment, this order being the lowest
integer $j\geq \lambda-1$.  This is reflected by the phase
transition scenario. For low values of $\lambda \leq 3$ the
system remains ordered for any finite temperature, whereas for
$\lambda > 3$ a finite temperature, order-disorder phase transition
occurs. Moreover, critical exponents that govern a second-order
phase transition in a scale-free network attain their usual
mean-field values for high $\lambda > 5$ and demonstrate non-trivial
$\lambda$-dependence in the region $3 < \lambda < 5$. Logarithmic
corrections to scaling laws appear at $\lambda=5$: this resembles
phenomena that occur at marginal space or order parameter
dimensions in bulk systems \cite{logarithmic}.

A starting point for our analysis will be the expression for the free
energy of a system with scalar order parameter on a scale-free
network. To be specific, from now on we will consider ferromagnetic
ordering and the spontaneous magnetization $M$ as the order parameter
with the conjugate magnetic field $H$. In this case the corresponding microscopic
degrees of freedom are the Ising spins. However,
generalization to models of more complicated symmetry is
straightforward \cite{Igloi02,Palchykov09}. Due to the fact that
the type of  networks under discussion are assumed to have a local tree-like structure, the
mean-field approximation is asymptotically exact in the sense that
thermal fluctuations can be neglected. This leads to a form of the
free energy also found by other techniques. The lowest order
contributions to the singular part of the Helmholtz free energy in the
vicinity of $T_c$ are  \cite{Leone02,Dorogovtsev02,Goltsev03}
\begin{eqnarray} \label{2.1}
F(M,T) &=&  \frac{a}{2}(T-T_c)M^2 + \frac{b}{4}M^4, \hspace{1em}
\lambda>5, \\ \label{2.2} F(M,T) &=&  \frac{a}{2}(T-T_c)M^2 +
\frac{b}{4}M^{\lambda-1}, \hspace{1em} 3 < \lambda < 5. \quad
\end{eqnarray}
The parameters $a,b>0$ and the critical temperature $T_c$ are
$\lambda$-dependent. This dependence can be made explicit using
microscopic approaches \cite{Leone02,Dorogovtsev02,Igloi02} or may be
postulated in a Landau-like approach \cite{Goltsev03,Palchykov09}.
For the subsequent analysis we will absorb the parameters into
the dimensions of the corresponding observables, passing to dimensionless
quantities,
\begin{eqnarray} \label{2.3}
f(m,\tau) &=&  \pm \frac{\tau}{2}m^2 + \frac{1}{4} m^4, \hspace{1em}
\lambda>5, \\ \label{2.4} f(m,\tau) &=& \pm \frac{\tau}{2} m^2 +
\frac{1}{4}m^{\lambda-1}, \hspace{1em} 3 < \lambda < 5,
\end{eqnarray}
with obvious relations between dimension-dependent and dimensionless
variables,
\begin{equation}\label{2.5}
m=M/M_0, \hspace{1em} \tau=|T-T_c|/T_c, \hspace{1em} f=F/F_0,
\end{equation}
where $M_0^{\lambda-3}=aT_c/b$ and $F_0=aT_cM_0^2$. Since $\tau$
measures the absolute distance to the critical point, the free
energy has two branches, corresponding to signs '$+$' and '$-$' in
Eqs. (\ref{2.3}) and (\ref{2.4}) for $T>T_c$ and $T<T_c$,
respectively. It is easy to verify, that a system with the free
energy (\ref{2.3}), (\ref{2.4}) possesses a second-order phase
transition at $\tau=0$. Here, we employ the standard notation for
critical exponents governing the temperature- and field-dependencies
of the thermodynamic functions. For $h=0$ and $T\rightarrow
T_c^\pm$, these are
\begin{eqnarray}
 c_h\simeq A^\pm  \tau^{-\alpha}, \quad
 \chi_T\simeq \Gamma^\pm \tau^{-\gamma}, \quad m_T\simeq B_T^\pm \tau^{-\omega}
 \label{notn1}
\end{eqnarray}
while for $T\rightarrow T_c^-$ one also has
\begin{equation}
 m\simeq B \tau^\beta.
 \label{notn2}
  \end{equation}
On the other hand, for $\tau=0$, the standard definitions are
\begin{eqnarray}
c_h & \sim    & A_c h^{-\alpha_c}, \quad
h    \simeq  D_c m |m|^{\delta-1}, \quad  \label{notn3}\\
\chi_T & \simeq & \Gamma_c h^{-\gamma_c}, \quad
m_T\simeq B_T^c h^{-\omega_c}.
 \label{notn4}
 \end{eqnarray}
(See section \ref{IIc} for the definition of the magnetocaloric
coefficient $m_T$.) The values of these critical exponents are
summarized in Table~\ref{tab1}
\cite{Leone02,Dorogovtsev02,Goltsev03,Igloi02,Palchykov10}. It is
worth noting here that one way to derive the listed exponents is to
consider the na\"ive dimensions of different terms in the Landau
free energy, similar to the standard field theoretical procedure
(see e.g. \cite{Amit}).
\begin{table}[t]
\begin{center}
\tabcolsep=0.5em
\begin{tabular}{ccccccccc}
\hline
             &$\alpha$&$\beta$&$\gamma$&$\delta$& $\omega$ & $\alpha_c$&$\gamma_c$ & $\omega_c$ \\\hline
  $\lambda\geq5$  &$0$&$1/2$&$1$&$3$&1/2 &$0$&$2/3$ &  1/3 \\
  $3<\lambda<5$&$\frac{\lambda-5}{\lambda-3}$&$\frac{1}{\lambda-3}$&$1$&$\lambda-2$& $\frac{\lambda-4}{\lambda-3}$ &
  $\frac{\lambda-5}{\lambda-2}$&$\frac{\lambda-3}{\lambda-2}$ & $\frac{\lambda-4}{\lambda-2}$
   \\
\hline
\end{tabular}
\end{center}
\caption{\label{tab1} Critical exponents governing temperature
and field dependencies of the thermodynamic
quantities for different values of $\lambda$.  }
\end{table}
With the values of critical exponents to hand, one can rewrite the
singular part of the Helmholtz potential in the form of a generalized
homogeneous function (\ref{1.2}) \cite{Stanley_Domb}:
\begin{eqnarray} \label{2.6}
f(m,\tau) & = & \tau^2 f_\pm (x), \hspace{1em} x= m/\tau^{\frac{1}{2}},
\hspace{1em} \lambda>5,
\\ \label{2.7} f(m,\tau) & =  & \tau^{\frac{\lambda-1}{\lambda-3}} f_\pm (x), \hspace{1em}
x= m/\tau^{\frac{1}{\lambda-3}}, \hspace{1em} 3 < \lambda < 5, \quad \quad
\end{eqnarray}
where the free energy scaling functions are given by
\cite{Palchykov10}
\begin{eqnarray} \label{2.8}
f_\pm (x) &=& \pm \frac{1}{2} x^2 + \frac{1}{4} x^4, \hspace{1em}
\lambda>5,
\\ \label{2.9}
 f_\pm (x) &=& \pm \frac{1}{2} x^2 +
\frac{1}{4}x^{\lambda-1}, \hspace{1em} 3 < \lambda < 5.
\end{eqnarray}
Assuming that the Helmholtz potential is a complete differential
\begin{equation} \label{2.10}
dF= -S dT + H d M
\end{equation}
one can further proceed with an analysis based on the magnetic
form of the equation of state,
\begin{equation} \label{2.11}
H = \frac{\partial F}{\partial M}\Big |_T
\end{equation}
or entropic form of the equation of state  \cite{note2},
\begin{equation} \label{2.12}
S = -\frac{\partial F}{\partial T}\Big |_M .
\end{equation}
As we have noted in the introduction, the scaling functions for the magnetic
equation of state (both in the Widom-Griffith
\cite{scaling_functions} and Stanley-Hankey \cite{Hankey72} forms)
and isothermal susceptibility have recently been  reported elsewhere
\cite{Palchykov10}. Here, we will proceed by analyzing the entropic
equation of state (\ref{2.12}) and heat capacity scaling functions.

In terms of dimensionless variables Eqs. (\ref{2.11}), (\ref{2.12}) take on
the form
\begin{equation} \label{2.13}
h (m,\tau) = \frac{\partial f(m,\tau)}{\partial m}\Big |_\tau,
\hspace{1em} s (m,\tau) = \mp \frac{\partial f(m,\tau)}{\partial
\tau}\Big |_m
\end{equation}
with field $h$ and entropy $s$ measured in units of $F_0/M_0$ and
$F_0/T_c$ correspondingly. As before, and throughout, the index
$\pm$ refers to temperatures above and below the critical point
$T_c$.

Since the free energy (\ref{2.1}), (\ref{2.2}) is explicitly a linear
function of $\tau$, one obtains the usual mean field result for the
 heat capacity at constant magnetization:
\begin{equation} \label{2.14}
C_M = T \frac{\partial S}{\partial T}\Big |_M=0 \, .
\end{equation}
To find the dimensionless constant-magnetic-field heat capacity
\cite{note2},
\begin{equation} \label{2.15}
c_h = \pm \frac{T}{T_c} \frac{\partial s(\tau,m)}{\partial \tau}\Big
|_h = (\tau \pm 1) \frac{\partial s(\tau,m)}{\partial \tau}\Big |_h,
\end{equation}
one can consider the entropy as a function of magnetic field and
temperature $s(\tau,m (\tau,h))$ which leads to
\begin{equation} \label{2.16}
c_h =  (\tau \pm 1)\Big [ \frac{\partial s}{\partial \tau}\Big |_m +
\frac{\partial s}{\partial m}\Big |_\tau \frac{\partial m
(\tau,h)}{\partial \tau}\Big |_h \Big ].
\end{equation}
Noting from (\ref{2.1}), (\ref{2.2}), that $\partial s/\partial m
|_\tau=-m$ and $\partial s/ \partial \tau |_m=0$ one finally arrives
at the  expression for the heat capacity,
\begin{equation} \label{2.17}
c_h = (1 \pm \tau) C_h (\tau,m)
\end{equation}
with function $C_h$ given by
\begin{equation} \label{2.18}
C_h (\tau,m) = \mp m \frac{\partial m (\tau,h)}{\partial \tau}\Big
|_h.
\end{equation}

Let us now consider separately the cases of fast ($\lambda>5$) and slower ($3<
\lambda < 5 $) decay of the node degree distribution (\ref{1.1}).

\subsection{$\lambda>5$ \label{IIa}}
The free energy (\ref{2.1}) leads to the  expression for
the entropy,
\begin{equation}\label{2.19}
s (\tau,m)= -\frac{m^2}{2},
\end{equation}
which can be easily recast in a scaling form
\begin{equation}\label{2.20}
s (\tau,m)= \tau {\cal S} (x),
\end{equation}
where the scaling variable $x=m/\tau^\beta=m/\tau^{1/2}$ and the entropy
scaling function ${\cal S}(x)$ is
\begin{equation}\label{2.21}
{\cal S}(x) = -\frac{x^2}{2}.
\end{equation}
 To obtain the heat capacity (\ref{2.16}) we first write the magnetic
 equation of state (\ref{2.13})
\begin{equation}\label{2.22}
h=\pm \tau m + m^3
\end{equation}
and differentiate it with respect to $\tau$ to obtain:
\begin{equation}\label{2.23}
\frac{\partial m}{\partial \tau} \Big |_h = \frac{\mp m}{\pm \tau +
3 m^2}.
\end{equation}
Substituting this into (\ref{2.18}) leads to the representation of
$C_h$ in the form of a generalized homogeneous function,
\begin{equation} \label{2.24}
C_h (\tau,m) = {\cal C_\pm}(x),
\end{equation}
with the scaling variable $x$ defined above and the heat capacity
scaling function
\begin{equation} \label{2.25}
{\cal C}_\pm(x) = \frac{x^2}{3x^2\pm 1}.
\end{equation}
 Note, that  in (\ref{2.24}) the heat capacity exponent
vanishes, $\alpha=0$.

\subsection{$3<\lambda<5$ \label{IIb}}

A particular feature of  the entropy of a system on a scale-free
network is that its dependence on magnetization both for
$3<\lambda<5$ and for $\lambda>5$ is given by Eq. (\ref{2.19}).
In terms of the scaling function for $3 < \lambda < 5$ it
reads
\begin{equation}\label{2.26}
s (\tau,x)= \tau^{2/(\lambda-3)} {\cal S} (x),
\end{equation}
where the entropy scaling function does not change and is given by
Eq. (\ref{2.21}). The power of $\tau$ is equal to $1-\alpha$ and the
scaling variable is now
\begin{equation}\label{2.27}
x\equiv m/\tau^\beta=m/\tau^{1/(\lambda-3)}.
\end{equation}
However, the magnetic equation of state (\ref{2.13}) for the
Helmholtz function (\ref{2.4}) becomes $\lambda$-dependent:
\begin{equation}\label{2.28}
h=\pm \tau m + \frac{\lambda-1}{4} m^{\lambda-2}.
\end{equation}
As in the previous subsection we obtain from this the derivative
$\partial m/\partial \tau |_h$, and by substitution into Eq.(\ref{2.18})
we arrive at the representation of $C_h$ in the form of
the generalized homogeneous function
\begin{equation} \label{2.29}
C_h (\tau,m) = \tau^{\frac{5-\lambda}{\lambda-3}} {\cal C_\pm}(x)
\end{equation}
where the scaling variable $x$ is given by (\ref{2.27}) and the heat
capacity scaling function attains a non-trivial $\lambda$-dependence
\begin{equation} \label{2.30}
{\cal C}_\pm(x) =
\frac{x^2}{\frac{(\lambda-1)(\lambda-2)}{4}x^{\lambda-3}\pm 1}.
\end{equation}

Note that on the basis of the  scaling functions
${\cal S}(x)$ in Eq.(\ref{2.21}) and ${\cal C}_\pm(x)$ in Eqs.(\ref{2.25}) and (\ref{2.30}),
 one easily obtains the corresponding scaling functions
with respect to the rescaled magnetic field
\begin{equation} \label{2.31}
y\equiv h/\tau^{\beta\delta}.
\end{equation}
The connection between the variables $x$ and $y$ results from
the magnetic equations of state (\ref{2.22}) and (\ref{2.28}), and is
given by
\begin{eqnarray}\label{2.32}
 y&=&\pm x + x^3, \hspace{4em} \lambda>5, \\ \label{2.33}
 y&=&\pm x + \frac{\lambda-1}{4} x^{\lambda-2}, \hspace{1em} 3 < \lambda < 5.
\end{eqnarray}
Solving the above equations with respect to $x$ and substituting the
result $x(y)$ into the functions ${\cal S}(x)$ and ${\cal C}_\pm(x)$ leads
to the scaling functions ${\cal S}(y)$ and ${\cal C}_\pm(y)$.
The behavior of the above scaling functions will be
analyzed in the next section. These functions together with the
scaling functions for the magnetic equation of state
$h=\tau^{\beta\delta}H_\pm(m/\tau^\beta)$ and isothermal
susceptibility $\chi_T=\tau^{-\gamma}\chi_\pm(m/\tau^{\beta})$
\cite{Palchykov10} are summarized in table \ref{tab2}.
\begin{table}[ht]
\begin{center}
\tabcolsep=1em
\begin{tabular}{ccc}
\hline
             & $3 < \lambda < 5$ & $\lambda>5$ \\\hline
$f_\pm$  &  $\pm x^2/2  + x^{\lambda-1}/4$ & $\pm x^2/2 + x^4/4$  \\
$H_\pm$  & $\frac{\lambda-1}{4}x^{\lambda-2} \, \pm \, x$  &  $x^3\, \pm \, x$ \\
$\chi_\pm$  & $\frac{1}{(\lambda-1)(\lambda-2)x^{\lambda-3}/4\, \pm \, 1}$ &  $\frac{1}{3x^2\, \pm \, 1}$ \\
${\cal S}$  &  $-x^2/2$ & $- x^2/2$  \\
${\cal C_\pm}$  & $\frac{x^2}{(\lambda-1)(\lambda-2)x^{\lambda-3}/4\, \pm \, 1}$  &  $\frac{x^2}{3x^2\, \pm \, 1}$  \\
${\cal M}_\pm$  & $\frac{x}{(\lambda-1)(\lambda-2)x^{\lambda-3}/4\, \pm \, 1}$  &  $\frac{x}{3x^2\, \pm \, 1}$  \\
$A^+/A^-$  & 0 & 0  \\
$\Gamma^+/\Gamma^-$ & $\lambda-3$ & 2  \\
$R_\chi $  & 1 & 1   \\
$R_C$  & 0 & 0  \\
$R_A$  & $\frac{1}{\lambda-2}[\frac{4}{\lambda-1}]^{\frac{\lambda-5}{(\lambda-2)(\lambda-3)}}$  & 1/3  \\
\hline
\end{tabular}
\end{center}
\caption{\label{tab2} Scaling functions  and amplitude ratios near
the critical point in scale-free networks. The scaling variable is
$x=m/\tau^\beta$. The ratio $\Gamma^+/\Gamma^-$ is taken from Ref.
\cite{Palchykov09} and scaling functions $f_\pm$, $H_\pm$, and
$\chi_\pm$ follow from  Ref. \cite{Palchykov10}.}
\end{table}

\subsection{Isothermal magnetocaloric effect, $\lambda>3$ \label{IIc}}

Before we proceed with the discussion of the peculiarities of the entropic
equation of state and of the thermodynamic functions following from
it, let us introduce an additional observable -- the isothermal
magnetocaloric coefficient. It serves as a direct measure of the
heat released by the system due to the magnetocaloric effect upon
an isothermal increase of the magnetic field and is defined as (see e.g.
\cite{magnetocaloric})
\begin{equation}\label{2.34}
M_T=-T\frac{\partial M}{\partial T}\Big |_H \, .
\end{equation}
In contradistinction to the heat capacity, which often does not
diverge or is a weakly divergent quantity for many 3d systems, the
magnetocaloric coefficient is frequently strongly divergent at
second-order phase transitions \cite{Plackowski05,magnetocaloric}
and therefore it is instructive to analyze how this behavior is
modified by a scale-free network. Using Maxwell relations, $M_T$ can
be obtained both from the magnetic or from the entropic equations of
state, Eqs. (\ref{2.11}), (\ref{2.12}). Therefore an equivalent
representation to the one given in (\ref{2.34}) representation
reads:
\begin{equation}\label{2.35}
M_T=-T\frac{\partial S}{\partial H}\Big |_T \, .
\end{equation}
Analogous to the first equation in (\ref{2.5}), we define the
dimensionless isothermal magnetocaloric coefficient as
\begin{equation}
 m_T = \frac{M_T}{M_0}\,.
\end{equation}
From the above representation this is
\begin{equation} \label{2.36}
m_T =  -(1 \pm \tau ) \frac{\partial s (m,\tau)}{\partial m}\Big
|_\tau \frac{\partial m (\tau,h)}{\partial h}\Big |_\tau.
\end{equation}
Recognizing that the last term in (\ref{2.36}) is a dimensionless
isothermal susceptibility $\chi_T(\tau,m)$ and writing it in the
scaling form
\begin{equation} \label{2.37}
\chi_T=\tau^{-\gamma}\chi_\pm(x), \hspace{1em} x=m/\tau^\beta ,
\end{equation}
we arrive at the scaling representation for the
dimensionless isothermal magnetocaloric coefficient $m_T$
\begin{equation} \label{2.38}
m_T=(1\pm \tau) \tau^{-\omega}{\cal M}_\pm(x),
\end{equation}
with the scaling function
\begin{equation} \label{2.39}
{\cal M}_\pm(x)=x\chi_\pm(x),
\end{equation}
and a scaling relation for the isothermal magnetocaloric coefficient
critical exponent $\omega$,
\begin{equation} \label{2.40}
\omega = 1 - \beta \, .
\end{equation}
While the equality (\ref{2.40}) is a general one and directly follows
from the scaling form of Eq. (\ref{2.36}), the relation (\ref{2.39})
between functions ${\cal M}_\pm(x)$ and $\chi_\pm(x)$ holds only for
systems where the entropy scaling function has the simple
representation (\ref{2.21}).

As noticed above, another way to obtain $m_T$ is to start from the
magnetic equation of state using the representation (\ref{2.34}). Then
one obtains
\begin{equation} \label{2.41}
m_T= \mp(1\pm\tau) \frac{\partial m (\tau,h)}{\partial \tau}\Big
|_h.
\end{equation}
Comparing this expression with the formulas obtained above for the
heat capacity (\ref{2.17}), (\ref{2.18}) one arrives at the
 relation between the scaling functions ${\cal M}_\pm(x)$
and ${\cal C}_\pm(x)$,
\begin{equation} \label{2.42}
{\cal C}_\pm(x) = x {\cal M}_\pm(x),
\end{equation}
which in particular  leads to (c.f. (\ref{2.39}))
\begin{equation} \label{2.43}
{\cal C}_\pm(x) = x^2 \chi_\pm(x).
\end{equation}

The scaling function ${\cal M}_\pm(x)$ defined  above is displayed for different
ranges of the values of $\lambda$ in table \ref{tab2}. For the critical
exponents $\omega$ we get
\begin{equation}\label{2.44}
\omega=\frac{\lambda-4}{\lambda-3}, \hspace{1em} 3<\lambda<5;
\hspace{1em} \omega=1/2, \hspace{1em} \lambda>5.
\end{equation}
It is easy to find the scaling relation for the critical exponent
$\omega_c$ that governs the field dependence of $m_T(\tau=0,h)$
\cite{Plackowski05}:
\begin{equation}\label{2.45}
\omega_c=\frac{1-\beta}{\beta\delta}\, .
\end{equation}
The values of this exponent read
\begin{equation}\label{2.46}
\omega_c=\frac{\lambda-4}{\lambda-2}, \hspace{1em} 3<\lambda<5;
\hspace{1em} \omega_c=1/3, \hspace{1em} \lambda>5.
\end{equation}
Thus while $c_h$ does not diverge ($\alpha <0$) for the entire range
$3 < \lambda <5$, $m_T$ is divergent ($\omega >0$) over half that
range  $4<\lambda <5$, and is a better locator of the phase
transition there. The above calculated exponents $\omega$,
$\omega_c$ are displayed together with other exponents in the
comprehensive table \ref{tab1}, which presents a summary of the data
concerning the temperature and field behavior of different
thermodynamic quantities in the vicinity of the critical point for
different values of $\lambda$. In the course of the analysis of
different types of critical phenomena in scale-free networks
\cite{Dorogovtsev08}, it has been revealed that the onset of
divergencies of moments of the node-degree distribution function
$P(k)$, Eq. (\ref{1.1}) relate to changes in the scaling scenario of
these systems. As one can see from (\ref{2.44}), (\ref{2.46}) the
exponents $\omega$ and $\omega_c$ change their sign to become
negative for $\lambda<4$: $m_T$ is not divergent at the critical
point any more in the region $3 < \lambda < 4$. Therefore, along
with the two marginal values of $\lambda=5$ and $\lambda=3$ which
correspond to the divergencies of $\langle k^4 \rangle$ and $\langle
k^2 \rangle$ and define the 'window' of non-trivial critical
behavior on a scale-free network. The divergence of the third moment
of the node-degree distribution $\langle k^3 \rangle$ leads to a
qualitative change in the critical behavior of the isothermal
magnetocaloric coefficient.

\section{Discussion \label{III}}

As one can see from table \ref{tab1}, the heat capacity exponent
$\alpha$ is negative in the region $3 < \lambda < 5$ where a
non-trivial $\alpha(\lambda)$ dependence is observed. This means
that the singular part of the heat capacity $c_h$ vanishes at $T_c$. Taken
that $c_h$ vanishes also at $T=0$ and that it is a positive smooth
function of $T$ in between, one concludes that it has
a maximum at some temperature $T_0$, where $0 < T_0 < T_c$ for any $3< \lambda <5$. Therefore
the energy fluctuations are maximal at $T_0$ (see \cite{Palchykov09}
for more details). Such behavior is a generic feature of systems
with $\alpha<0$, other examples include  the three dimensional
Heisenberg and planar magnets \cite{Heisenberg}, liquid helium
\cite{Helium}, and disordered uniaxial magnets \cite{RIM}. From Eqs.(\ref{2.25}) and (\ref{2.30})
one finds that $c_h(T>T_c,h=0)=0$ for
any $\lambda$. This leads to the following amplitude ratio which holds for all
$\lambda>3$:
\begin{equation} \label{3.1}
A^+/A^-=0.
\end{equation} Amplitude ratios are known to be universal along with
the scaling functions and critical exponents (see e.g.
\cite{Privman91}). It is appropriate to adduce here how do these
ratios change for systems on scale-free networks. The results
are summarized in the lower part of table \ref{tab2}. Besides than
the heat capacity amplitude ratio (\ref{3.1}), the isothermal
magnetic-susceptibility amplitude ratio appears to be
$\lambda$-dependent for $3 < \lambda < 5$:
$\Gamma^+/\Gamma^-=\lambda-3$ \cite{Palchykov09}. Using the
expressions (\ref{2.22}), (\ref{2.25}), (\ref{2.28}), (\ref{2.30})
it is straightforward to find for the other amplitudes for
$\lambda>5$
\begin{equation}\label{3.3}
B=D_c=1, \hspace{2em} A_c=1/3,
\end{equation}
and for $3 < \lambda < 5$,
\begin{eqnarray}\nonumber
B&=&\Big(\frac{4}{\lambda-1}\Big )^{1/(\lambda-3)}, \hspace{2em}
D_c=\frac{\lambda-1}{4} \, , \\ \label{3.4} A_c&=&
\frac{1}{\lambda-2} \Big (\frac{4}{\lambda-1}\Big )^{3/(\lambda-2)}
\, .
\end{eqnarray}
Now, defining three more amplitude ratios by
\cite{Barmatz75,Aharony76,Privman91}
\begin{eqnarray} \label{3.5}
R_{\chi}&=&\Gamma^+D_cB^{\delta-1}, \\ \label{3.6}
R_c&=&A^+\Gamma^+/B^2, \\ \label{3.7}
R_A&=&A_cD_c^{-(1+\alpha_c)}B^{-2/\beta},
\end{eqnarray}
and substituting into these ratios the amplitudes (\ref{3.3}) and
(\ref{3.4}), we arrive at their values for the scale-free network, as
listed in table \ref{tab2}.

\begin{figure}[h!]
 \begin{center}
 \includegraphics[width=9cm]{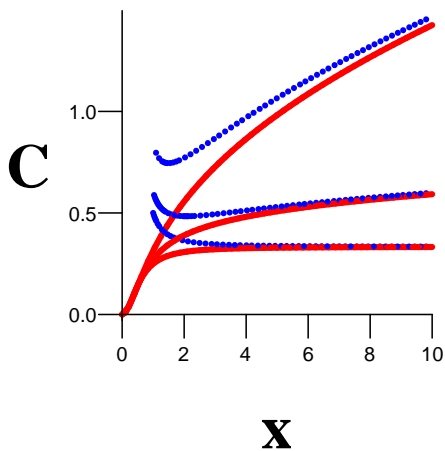}
 \vspace{-7ex}
 \caption{\label{fig1} Heat capacity scaling functions
${\cal C}_-(x)$ (dotted curves, blue online) and ${\cal C}_+(x)$
(solid curves, red online) as functions of the scaling variable
$x=m/\tau^\beta$ at $\lambda>5$, $\lambda=4.8$, and $\lambda=4.5$
(lower, middle and upper pairs of curves, respectively).}
\end{center}
\end{figure}

Let us concentrate now on the scaling functions. As noted
in section \ref{II}, the entropy scaling function ${\cal S}_\pm(x)$
which in the usual Landau theory is given by Eq. (\ref{2.21}) keeps
its form also in the case of scale-free networks with $3< \lambda <
5$. However, the constant-magnetic-field heat capacity scaling
function ${\cal C}_\pm$ essentially changes in this region. In
Fig. \ref{fig1} we plot ${\cal C}_\pm$  as a function of the scaling
variable $x=m/\tau^\beta$ for different values of $\lambda$. The
most striking feature in the behavior of the scaling function is
that its asymptotics for large $x$ change for $\lambda<5$.
Indeed, for $\lambda > 5$ the asymptotical value is given by a
constant, ${\cal C}_\pm(x\to \infty)= 1/3$, whereas in the range $3< \lambda
<5$ the function behaves as a power law,
\begin{equation}\label{3.8}
{\cal C}_\pm(x\to \infty) = \frac{4}{(\lambda-1)(\lambda-2)}\,
x^{5-\lambda}.
\end{equation}
In turn, this is reflected in the development of a minimum in the
$C_-$ branch of the function as $\lambda$ decreases (see the
figure).

Another particular feature of the plots of Fig. \ref{fig1} is
inherent to the presentation of the scaling plots in the ${\cal
C}_-$--$x$-plane and is connected to the presence of a pole in
${\cal C}_\pm(x)$ for small $x$. As one sees immediately from
Eqs. (\ref{2.25}) and (\ref{2.30}), this pole occurs at $x=1/\sqrt{3}$
and $x=\big(4/[(\lambda-1)(\lambda-2)]\big )^{1/(\lambda-3)}$ for
$\lambda=5$ and $3< \lambda < 5$, correspondingly. However, the
physical values of the scaling variable $x$ do not cover the region
where the pole occurs. Indeed, from the magnetic equations
of state (\ref{2.22}), (\ref{2.28}) one may obtain the solutions for the magnetization
at zero magnetic field $m(\tau,h=0)$. Taking that a non-zero magnetic
field must increase the value of $m$ one arrives at the
following minimal values of the scaling variable $x$:
\begin{eqnarray}\label{3.9}
x_{\rm min}&=&1, \hspace{1em} \lambda>5, \\ \label{3.10}
 x_{\rm min}&=&\Big ( \frac{4}{\lambda-1} \Big )^{1/(\lambda-3)} , \hspace{1em} 3<\lambda<5.
\end{eqnarray}
Therefore, the curves for the scaling function ${\cal C}_-$ in Fig. \ref{fig1}
originate at the corresponding minimal values of $x$  defined by the relations
(\ref{3.9}), (\ref{3.10}).

\begin{figure}[th]
 \begin{center}
\includegraphics[width=9cm]{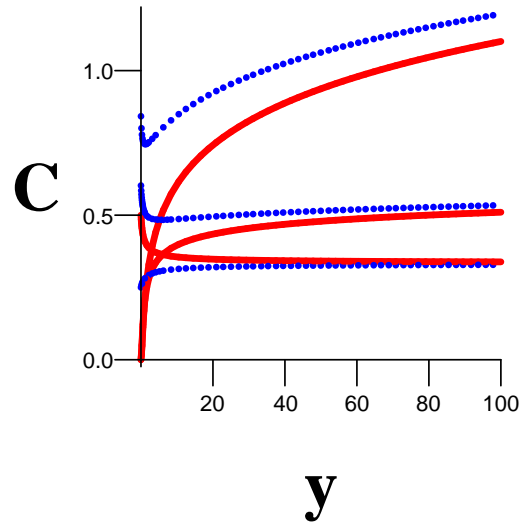}
\vspace{-7ex}
 \caption{\label{fig2} Heat capacity scaling functions
${\cal C}_-(y)$ (dotted curves, blue online) and ${\cal C}_+(y)$
(solid curves, red online) as functions of the scaling variable
$y=h/\tau^\beta\delta$ at $\lambda > 5$, $\lambda=4.8$, and
$\lambda=4.5$ (lower, middle and upper pairs of curves,
respectively).}
\end{center}
\end{figure}

In turn, as explained in section \ref{II}, one may re-express the
scaling function ${\cal C}_\pm$ in terms of the scaled magnetic
field $y$ using Eq. (\ref{2.31}). Corresponding  plots for the scaling function ${\cal C}_\pm(y)$ in this variable
are given in Fig. \ref{fig2} for different values of $\lambda$.
Again, one observes a change in the asymptotics of the scaling
function: instead of a constant at $\lambda>5$, for $3<\lambda<5$
the asymptotic functional dependence is given by
\begin{equation}\label{3.11}
{\cal C}_\pm(y\to \infty) = \frac{1}{(\lambda-2)} \Big
(\frac{4}{(\lambda-1)}\Big )^{\frac{3}{\lambda-2}}\,
y^{\frac{5-\lambda}{\lambda-2}}.
\end{equation}

\begin{figure}[t]
 \begin{center}
\includegraphics[width=9cm]{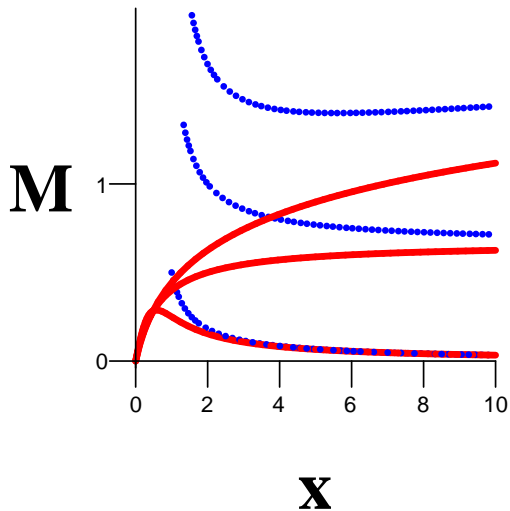}
\vspace{-7ex}
 \caption{\label{fig3}  Scaling functions fo the
isothermal magnetocaloric coefficient.
 ${\cal M}_-(x)$ (dotted curves, blue online) and ${\cal
M}_+(x)$ (solid curves, red online) as functions of the scaling
variable $x=m/\tau^\beta$ at $\lambda > 5$, $\lambda=4$, and
$\lambda=3.8$ (lower, middle and upper pairs of curves,
respectively). The decay which is observed for $\lambda>4$ switches
to power law growth for $\lambda>4$. The functions approach the
constant value ${\cal M}_\pm=2/3$ for $\lambda=4$. }
\end{center}
\end{figure}

\begin{figure}[t]
 \begin{center}
\includegraphics[width=9cm]{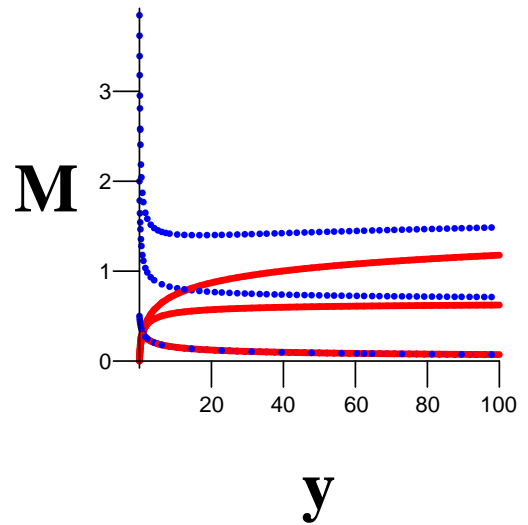}
\vspace{-7ex}
 \caption{\label{fig4} Isothermal magnetocaloric
coefficient scaling functions ${\cal M}_-(y)$ (dotted curves, blue
online) and ${\cal M}_+(y)$ (solid curves, red online) as functions
of the scaling variable $y=m/\tau^{\beta\delta}$ at $\lambda > 5$,
$\lambda=4$, and $\lambda=3.8$ (lower, middle and upper pairs of
curves, respectively).}
\end{center}
\end{figure}

In figures \ref{fig3} and \ref{fig4} we give the plots of the  scaling functions
${\cal M}_\pm$ for the
isothermal magnetocaloric coefficient in the scaling variables $x=m/\tau^\beta$ and
$y=h/\tau^{\beta\delta}$, respectively. As  discussed at
the end of the previous section, $m_T$ changes its behavior at
$\lambda=4$. This feature is also reflected in the behavior of the
scaling functions: their asymptotics  change at $\lambda=4$. Indeed,
for $\lambda>5$ the function decays as ${\cal M}_\pm(x\to
\infty)\sim 1/3x$ whereas from the asymptotic behavior in the region
$3 < \lambda < 5$,
\begin{equation}\label{3.12}
{\cal M}_\pm(x\to \infty) = \frac{4}{(\lambda-1)(\lambda-2)}\,
x^{4-\lambda},
\end{equation}
one concludes, that for $\lambda<4$ the power law decay
switches to a power law growth, while ${\cal M}_\pm(x\to \infty)=const$ for the
marginal value $\lambda=4$. The corresponding asymptotics in the
variables $y$ is of the form
\begin{eqnarray}\label{3.13}
{\cal M}_\pm(y\to \infty) &=& \frac{1}{3y^{1/3}} \, ,  \hspace{1em} \lambda>5, \\
\nonumber {\cal M}_\pm(y\to \infty) &=& \frac{1}{(\lambda-2)} \Big
(\frac{4}{(\lambda-1)}\Big )^{\frac{2}{\lambda-2}}\,
y^{\frac{4-\lambda}{\lambda-2}} \, , \\ \nonumber && \hspace{3em} 3
\, < \, \lambda \, < \, 5 \, .
\end{eqnarray}

\section{Conclusions \label{IV}}

Usually the universality of critical phenomena is attributed to the
presence of only a few relevant global parameters. Different systems
that share the same values of these global variables manifest the
same criticality. A classical example is given by the famous 3d
Ising model universality class that is inherent to the critical
behavior of such differing systems as uniaxial magnets, simple
fluids or binary alloys. The critical behavior in all these systems
is quantitatively described by the same values of the critical
exponents, amplitude ratios, and by the same form of the scaling
functions. As this paper demonstrates, in particular the usual
`Euclidean space' understanding of universality of critical
phenomena breaks down if the critical behavior occurs on a
scale-free network. The presence of high-degree vertices (hubs) may
lead to substantial changes in ordering processes. The parameter
that controls the `importance' of the hubs is the node-degree
distribution exponent $\lambda$, Eq. (\ref{1.1}), and it is this
parameter that plays the role of a global variable as far as the
critical behavior is considered.

In particular, the universal quantities that govern criticality
become $\lambda$-dependent for small enough $\lambda$ and in this
way the network structure is felt. However, the presence of
magnetization is necessary to `feel' the network structure. To give
an example, the structure matters for $T<T_c$ at any $h$ and for
$T>T_c$ for $h \neq 0$ (c.f. that amplitude $\Gamma_-$ is
$\lambda$-dependent, whereas $\Gamma_+$ is not). Another interesting
observation is that the fluctuation in network structure  only
enters via the magnetization and since the entropy $S$ is measured
at constant magnetization, it is given by a usual Landau-like mean
field value for any $\lambda>3$. This makes a difference between the
global parameters $\lambda$ and dimension $d$ regarding its
influence of the fluctuations on calculating the singular behaviour
of the physical quantities: no renormalization-group calculation is
necessary.

In this paper we completed the quantitative description of critical
behavior in scale-free networks by calculating the entropic equation
of state, the resulting scaling functions as well as the universal amplitude ratios.
The corresponding formulas, together with other data for the critical
exponents and amplitudes are summarized in Tables \ref{tab1} and
\ref{tab2}.  They constitute a comprehensive list of observables
that describe the scaling and characterize the criticality in scale-free
networks.

The starting point for our study was the asymptotic form of the
free energy in the vicinity of a critical point in a scale-free
network, Eqs. (\ref{2.1}), (\ref{2.2}). The validity of this
expression has been proven at different levels of rigor
using microscopic approaches based on the recursion relations
\cite{Dorogovtsev02} the replica method \cite{Leone02} or
phenomenological Landau approaches \cite{Goltsev03,Palchykov09} as well as
mean field theory \cite{Igloi02}. It is instructive to note here
that, because the networks under discussion have a local tree-like
structure, the mean-field approximation is asymptotically exact. One
of the consequences of this fact is that the values of the exponents do
not change if an $O(m)$-symmetrical order parameter is considered
(see e.g. \cite{Palchykov09}), as it is usual in the Landau theory.

\section*{Acknowledgement}

This work was supported by the Austrian Fonds zur F\"orderung der
wissenschaftlichen Forschung under Project No. P19583-N20 and by the
ARF Scheme of Coventry University.


\begin{thebibliography}{99}

\bibitem{Dorogovtsev08}  S.~N.~Dorogovtsev and A.~V.~Goltsev, Rev. Mod. Phys. {\bf 80}, 1275 (2008).

\bibitem{sociophysics}
S. Galam, Physica A {\bf 274}, 132 (1999); Int. J. Mod. Phys. C {\bf
19}, 409 (2008); K. Sznajd-Weron and J. Sznajd, {\em ibid.} {\bf
11}, 1157 (2000); K. Sznajd-Weron, Acta Phys. Pol. B {\bf 36}, 2537
(2005); D. Stauffer and S. Solomon, Eur. Phys. J. B {\bf 57}, 473
(2007); K. Ku{\l}akowski and M. Nawojczyk, e-print arXiv:0805.3886.

\bibitem{Tadic05}
B. Tadi\'c, K. Malarz, and K. Ku{\l}akowski, Phys. Rev. Lett. {\bf
94}, 137204 (2005).

\bibitem{Stanley_Domb} See, e.g., H.~E.~Stanley, {\em Introduction to
Phase Transitions and Critical Phenomena} (Oxford University Press,
Oxford, 1971); C.~Domb, {\em The Critical Point} (Taylor \& Francis,
London, 1996).

\bibitem{note1} If the Helmholtz potential $F(\tau,m)$
is a generalized homogeneous function, one can show \cite{Hankey72}
that the other thermodynamic potentials are also generalized
homogeneous functions.

\bibitem{Hankey72} A.~Hankey, H.~E.~Stanley, Phys. Rev. B {\bf 6},
3515 (1972).

\bibitem{scaling_functions} C.~Domb and D.~L.~Hunter, Proc. Phys.
Soc. {\bf 86}, 1147 (1965); R.~B.~Griffiths, Phys. Rev. {\bf 158},
176 (1967); A.~Z.~Patashinskii and V.~L.~Pokrovskii, Zh. Eksp. Teor.
Fiz. {\bf 50}, 439 (1966) [Sov. Phys. JETP {\bf 23}, 292 (1966)];
B.~Widom, J. Chem. Phys. {\bf 43}, 3898 (1965).

\bibitem{Privman91}
See e.g. V.~Privman, P.~C.~Hohenberg, and A.~Aharony, in: {\em Phase
Transitions and Critical Phenomena}  (Edited by C.~Domb,
J.~L.~Lebowitz), {\bf 14}, 1-134 (1991).

\bibitem{Leone02}
 M.~Leone, A.~V\'{a}zquez, A.~Vespignani, and R.~Zecchina, Eur. Phys. J. B {\bf 28}, 191
 (2002).

\bibitem{Dorogovtsev02}
 S.~N.~Dorogovtsev, A.~V.~Goltsev, and J.~F.~F.~Mendes, Phys. Rev. E {\bf 66}, 016104
 (2002).

\bibitem{Goltsev03} A.~V.~Goltsev, S.~N.~Dorogovtsev,  and J.~F.~F.~Mendes, Phys. Rev.
E {\bf 67}, 026123 (2003).

\bibitem{Igloi02}
F.~Igl\'oi and L.~Turban, Phys. Rev. E {\bf 66}, 036140 (2002).

\bibitem{Palchykov09}
V.~Palchykov, C.~von~Ferber, R.~Folk, and Yu.~Holovatch, Phys. Rev.
E {\bf 80}, 011108 (2009).

\bibitem{Palchykov10}
V.~Palchykov, C.~von~Ferber, R.~Folk, Yu.~Holovatch, and R. Kenna.
Phys. Rev. E {\bf 82}, 011145 (2010).

\bibitem{Amit} D.J. Amit. {\em Field Theory, the Renormalization
Group, and Critical Phenomena} (World Scientific, Singapore, 1989).

\bibitem{Krasnow73} R. Krasnow and H.E. Stanley. Phys. Rev. B
{\bf 8}, 332 (1973).

\bibitem{Simons74} D.S. Simons and M.B. Salamon. Phys. Rev. B
{\bf 10}, 4680 (1974).

\bibitem{Barmatz75} M. Barmatz, P.C. Hohenberg, and A. Kornblit. Phys. Rev. B
{\bf 12}, 1947 (1975).

\bibitem{Aharony76} A. Aharony and P.C. Hohenberg. Phys. Rev. B
{\bf 13}, 3081 (1976).

\bibitem{Salamon88} M.B. Salamon, S.E. Inderhees, J.P. Rice, B.G. Pazol,
D.M. Ginsberg, and N. Goldenfeld. Phys. Rev. B {\bf 38}, 885 (1988).

\bibitem{Stroka92} B. Stroka, J. Woznitza, E. Scheer, H. v. L\"ohneysen,
W. Park, and K. Fischer.  Z. Phys. B {\bf 89}, 39 (1992).

\bibitem{Plackowski05} T. Plackowski and D. Kaczorowski. Phys. Rev.
B {\bf 72}, 224407 (2005).

\bibitem{Franco09} V. Franco, A. Conde, J.W. Romero-Enrique, Y.I. Spichkin,
and V.I. Zverev, A.M. Tishin. J. Appl. Phys. {\bf 106}, 103911
(2009).

\bibitem{logarithmic} F. J. Wegner, in {\em Phase Transitions and
Critical Phenomena}, edited by C. Domb and M. S. Green (Academic
Press, London, 1976), Vol. VI, p. 8; R.~Kenna, D.~A.~Johnston, and
W.~Janke, Phys. Rev. Lett. {\bf 96}, 115701 (2006); Phys. Rev. Lett.
{\bf 97}, 155702 (2006), Phys. Rev. Lett. {\bf 97}, 169901(E)
(2006).

\bibitem{note2} Entropies at constant magnetic field and at
constant magnetization coincide.

\bibitem{magnetocaloric} V.K. Pecharsky, K.A. Gschneidner Jr.
J. Mag. Mag. Mat. {\bf 200}, 44 (1999); T. Plackowski, A. Junod, F.
Bouquet, I. Sheikin, Y. Wang, A. Je\.zowski, and K. Mattenberger.
Phys. Rev. B {\bf 67}, 184406 (2003); T. Plackowski, D. Kaczorowski,
and Z. Bukowski. Phys. Rev. B {\bf 72}, 184418 (2005).

\bibitem{Heisenberg} D. L. Connelly, J. S. Loomis, and D. E. Mapother.
Phys. Rev. B {\bf 3}, 924 (1971); F. L. Lederman, M. B. Salamon, and
L. W. Shacklette. Phys. Rev. B {\bf 9}, 2981 (1974).

\bibitem{Helium} J. A. Lipa, D. R. Swanson, T. C. P. Chui, and U. E.
Israelsson. Phys. Rev. Lett. {\bf 76}, 944 (1996).

\bibitem{RIM} R. Folk, Yu. Holovatch, and T. Yavors'kii.
Uspiekhi Fizichieskikh Nauk {\bf 173}, 175 (2003) [Physics-Uspiekhi,
{\bf 46}, 169 (2003)].

\end{thebibliography}
\end{document}